\documentclass[review]{elsarticle}

\usepackage{lineno,hyperref}
\modulolinenumbers[5]
\usepackage{color}
\usepackage{ulem}
\usepackage{tabularx,ragged2e,booktabs,caption,graphicx,epstopdf}

\journal{Carbon}

\newcommand{\hp}{H$_2$O$_2$}
\newcommand{\fen}{Fe(NO$_3$)$_3$}
\newcommand{\km}{KMnO$_4$}
\newcommand{\oa}{C$_2$H$_2$O$_4$}
\newcommand{\sod}{Na$_2$S$_2$O$_3$}
\newcommand{\sd}{sodium thiosulfate}
\newcommand{\ox}{oxalic acid}
\newcommand{\pp}{potassium permanganate}
\newcommand{\fn}{ferric nitrate}
\newcommand{\hy}{hydrogen peroxide}

\begin{document}

\begin{frontmatter}

\title{Redox agent enhanced chemical mechanical polishing of thin film diamond}

\author[1,3]{Soumen Mandal\corref{a}}
\ead{mandals2@cardiff.ac.uk}
\author[1,3]{Evan L. H. Thomas\corref{a}}
\ead{thomasel10@cardiff.ac.uk}
\author[1]{Laia Gines}
\author[4]{David Morgan}
\author[1]{Joshua Green}
\author[2]{Emmanuel B. Brousseau}
\author[1]{Oliver A. Williams\corref{a}}
\ead{williamso@cardiff.ac.uk}
\address[1]{School of Physics and Astronomy, Cardiff University, Cardiff, UK}
\address[2]{School of Engineering, Cardiff University, Cardiff, UK}
\address[4]{Cardiff Catalysis Institute, School of Chemistry, Cardiff University, Cardiff, UK }
\address[3]{These authors contributed equally to the work}
\cortext[a]{Corresponding authors}

\begin{abstract}
The chemical nature of the chemical mechanical polishing of diamond has been examined by adding various redox agents to the alkaline SF1 polishing slurry. Three oxidizing agents namely, \hy, \pp{} and \fn{}, and two reducing agents, \ox{} and \sd{}, were added to the SF1 slurry. Oxalic acid produced the fastest polishing rate while \hy{} had very little effect on polishing, probably due to its volatile nature. X-ray photoelectron spectroscopy (XPS) reveals little difference in the surface oxygen content on the polished samples using various slurries. This suggests that the addition of redox agents do not increase the density of oxygen containing species on the surface but accelerates the process of attachment and removal of Si or O atoms within the slurry particles to the diamond surface.
\end{abstract}


\end{frontmatter}


\section{Introduction}

Recent advances in chemical vapour deposition (CVD) have resulted in the growth of high quality, nanocrystalline diamond atop large area substrates \cite{williamsrev}.  With such films demonstrating properties rivalling many of those of single crystal diamond \cite{williams2010,angadi2006}, the potential applications of NCD range from the diverse areas of micro-electro-mechanical systems (MEMS) and surface acoustic wave (SAW) devices to optical coatings \cite{williamsrev,gaidarzhy2007}.  However, the large surface energy of diamond relative to the substrate coupled with the low sticking probability of methyl precursors typically ensures that a seeding step is required to realise coalesced thin films \cite{williamsrev}.  Upon subjecting to CVD, growth then proceeds atop the seeds resulting in the formation of isolated islands capable of growth both normal and parallel to the substrate \cite{jiang1994}.  After sufficient lateral growth these islands will coalesce and form columnar crystals in accordance with the van der Drift growth model.  Due to the differing growth rates of the diamond facets, competitive overgrowth of the columnar crystals will then occur leading to a surface roughness that evolves with films thickness \cite{smereka2005}.  This substantial roughness then acts to spoil the Q-factor in MEMS devices, disperse acoustic waves in SAW applications, and cause diffusion in optical applications \cite{pala2007, flannery2002, ergincan2012,malshe1999}. 

To minimise these losses the roughness can be reduced during growth, circumvented upon processing, or polished away post growth;  should samples be grown with either an increased CH$_4$/H$_2$ fraction or Ar/CH$_4$ based chemistries, the lack of etching from atomic hydrogen will result in the formation of distinct secondary-nuclei upon crystal facets and the termination of the existing growth \cite{williamsrev}.  This `re-nucleation' process will then limit the grain size and prevent competitive overgrowth, leading to a roughness independent of film thickness.  However the corresponding increase in grain boundary region will result in the incorporation of a larger non sp$^3$ carbon content, reducing the Young's Modulus, the optical transparency, and the thermal conductivity of the resulting films \cite{williams2010, angadi2006, nesladek1996}.  Futhermore, modification of the gas chemistry will not be able to produce samples with roughness \textless\ 5 nm RMS, prohibiting use in applications requiring tightly controlled deposition of subsequent layers \cite{iriarte2010}.  Alternatively, the typically smoother nucleation side of the NCD film can be utilised through either selective etching of the substrate or through bonding the upturned film to a support structure.  However, once again, the small surface/volume ratio of the initial stages of film growth leads to a significant non diamond content within grain boundaries and a corresponding degradation in film properties, while the bonding processes are non-trivial, often require tens of microns thick diamond, and are incompatible with many applications \cite{thomas2017, mortet2002, rodri2012}.  Therefore, an effective means of polishing NCD films needs to be developed to fully utilise the properties of thin film diamond.

Traditionally, the polishing of diamond has predominantly focused on removal through pressing the sample against a rotating cast iron or alumina scaife embedded with diamond grit \cite{hird2004}.  However, this process introduces subsurface damage which can propagate microns into the bulk, leading to an increase in dislocation defects \cite{malshe1999,pastewka2011, friel2009, schuelke2013}.  For samples in which the film thickness is far smaller than the wafer bow meanwhile, the large pressures (ca. 0.1 MPa) exerted during polishing against the rigid metal scaife will often shatter the substrate, or at best, lead to uneven polishing of the wafer \cite{ollison1999, thomas2014}.  Doing away with these issues, the recent adaption of the technique of chemical mechanical polishing (CMP) from the IC-industry to diamond has demonstrated the polishing of highly bowed NCD films to roughness values below 2 nm RMS over areas of 25 $\mu$m$^2$, without the risk of sub-surface damage or significant polishing introduced contamination \cite{thomas2014}.  Further developments of the technique have demonstrated the role of the abrasive size in polishing, highlighting the mechanical nature of the technique, while density function theory studies have focused on the dynamics of surface degradation by silica, demonstrating the chemical aspects of polishing  \cite{werrell2017,peguiron2016}.  In this work the addition of redox agents to the slurry is then explored in an attempt to improve the removal rate, and elucidate on the chemical nature of the polishing mechanism.  

\section{Experiments}

For the purposes of polishing a series of $\approx$350nm thick diamond films were grown atop silicon dioxide buffered silicon substrates using the process detailed within a previous study \cite{werrell2017}.  Atomic force microscopy (AFM) scans were then obtained post growth with a Park Systems Park XE-100, utilising a Tespa-V2 tip and operating in non-contact mode.  A roughness value of 24.5 $\pm$ 1.5 nm over an area of 25 $\mu$m$^2$ was exhibited for the as grown films after analysis of the scans with Parks System XEI data processing software and averaging over the series of films grown.

Polishing was performed with a Logitech Tribo polishing system equipped with a Suba-X polyester impregnated polyurethane felt.  The polishing slurries were prepared in-house by mixing various redox agents in Logitech supplied Syton SF1, an alkaline colloidal silica slurry.  As per manufacturer's datasheet, SF1 consists of 15-50\% silicon dioxide particles with 4-5\% glycol mixed in water. The pH of the unaltered slurry is between 9.2 and 10.1. Three different oxidisers were investigated, hydrogen peroxide (\hp), ferric nitrate (\fen), and potassium permangante (\km), along with two reducing agents, oxalic acid (\oa), and sodium thiosulfate (\sod).  The concentration of \hp{} in the slurry was 20\% by volume while the other solid agents had a concentration of around 2.5gm/l.  As the addition of \ox{} made the resulting slurry highly acidic, sodium hydroxide (NaOH) was added to the slurry to bring the pH to a value compatible with the polishing pad.  With pH having previously been demonstrated to have little effect on the polishing rate and the remaining slurries all exhibiting values within the operating range of the pad, their pH values were left unaltered \cite{werrell2017}.

Before polishing, the pads were conditioned for 30mins with a conditioning chuck which consists of nickel plate embedded with diamond grit and polishing slurry to facilitate the polishing and slurry distribution\cite{thomas2014, thomas2014sc}.  For polishing the films, both the pad and wafer carrier were rotated at 60rpm opposite to each other with carrier sweeping across the pad. Two samples were then polished for each slurry at a downward pressure of 2 psi and intervals of 1 hour, until the roughness value dropped below 2 nm RMS over 25$\mu$m$^2$ or for a maximum of 4 hours \cite{werrell2017, thomas2014}.  The bow in the wafer was crudely adjusted by applying a 20psi backing pressure on the wafer carrier. After polishing the films were cleaned with standard SC-1 process. In this process the polished wafer was dipped in a solution of 30\% H$_2$O$_2$, NH$_4$OH and DI H$_2$O in the ratio of 1:1:5 at 75 $^o$C for 10min and then rinsed with DI water followed by spin drying. After cleaning, three areas on each sample were scanned and the RMS roughness values were averaged over the three areas.

XPS experiments were conducted with a Thermo K-alpha+ spectrometer at a pressure of ~$3\times10^{-7}$ mbar under an argon environment to facilitate charge neutralisation. Using an Al K$\alpha$ radiation source (1486.6 eV) operating at 12 kV anode potential and 6 mA emission current, `broad' survey scans and `narrow' scans of the pertinent peaks were obtained at pass energies of 150 eV and 40 eV, respectively.  Analysis was performed with CasaXPS, with the binding energies calibrated to the bulk sp$^3$ component of the C1s peak, and the peak areas normalised through the use of atomic sensitivity factors \cite{Wagner1981}.  Finally, particle size analysis was performed with a Malvern Zetasizer Nano Z dynamic light scattering system.

\section{Results and Discussions}

\begin{table}
\centering
\begin{tabular}{ c c c }
\hline
Slurry & Polishing duration & Final roughness \\ 
& (hr) & (RMS nm) \\
\hline
SF1 & 4 & 2.7\\
SF1 + \hp (20\% v/v) & 4 & 2.3\\
SF1 + \fen (2.5 gm/l)& 3 & 1.9\\
SF1 + \km (2.5 gm/l)& 3 & 1.9\\
SF1 + \oa (2.5 gm/l)& 3 & 1.8\\
SF1 + \sod (2.5 gm/l)& 3 & 2.0
\end{tabular}
\caption{Duration of polishing and final roughness values achieved for the 6 polishing slurries.}
\label{tab1}
\end{table}

\begin{figure}
\centering
\includegraphics[height=0.8\textheight]{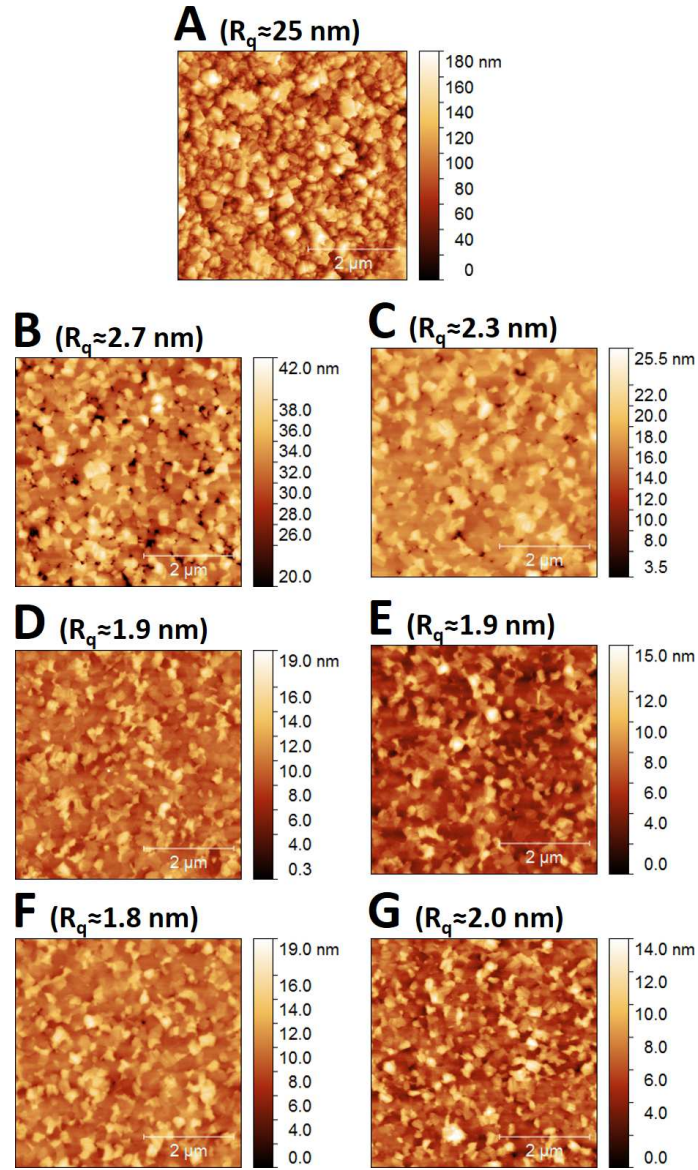}
\caption{A) AFM images of as grown film, and B)-G) final films polished with  SF1 + \hp, SF1 + \fen, SF1 + \km, SF1 + \oa, SF1 + \sod\ respectively with the final RMS roughness values indicated within brackets.} \label{figafm}
\end{figure}

Figure \ref{figafm} shows the AFM images of the as grown and SF1, SF1 + \hp, SF1 + \fen, SF1 + \km, SF1 + \oa, and SF1 + \sod\ polished samples in panels A and B-G, respectively. Meanwhile, table \ref{tab1} summarizes the polishing duration and average roughness values achieved for each slurry.  As visible within panel A, the as grown sample exhibits a variety of texture and a large maximum displacement of $\approx$180 nm as a result of competitive van der Drift type overgrowth.  Averaging the 3 areas scanned for each of the films within the growth series then yields a corresponding roughness of 24.5 $\pm$ 1.5 nm RMS over an area of 25 $\mu$m$^2$.  Upon polishing for 4 hours with SF1 a clear difference can be observed within figure \ref{figafm}B with the removal of the crystallite peaks and the formation of plateaus atop the grains, leading to a reduction in roughness to 2.7 $\pm$ 0.4 nm RMS.  Similarly, panel C shows the sample polished with the addition of \hp\ to the polishing slurry, with the additive leading to a minor reduction in the roughness to 2.3 $\pm$ 0.3 nm.  The samples polished for 3 hours with the addition of \fn, \pp, \ox, and \sd\ within panels D-G respectively however show a marked difference; polishing has progressed to the point at which neighbouring crystals intersect with each other, reducing the number of surface pits visible.  As a result, roughness values of 1.9 $\pm$ 0.1, 1.9 $\pm$ 0.1, 1.8 $\pm$ 0.1 and 2.0 $\pm$ 0.1 nm for \fn, \pp, \ox\, and \sd\ additives, respectively, are achieved.

\begin{figure}
\centering
\includegraphics[height=0.6\textwidth]{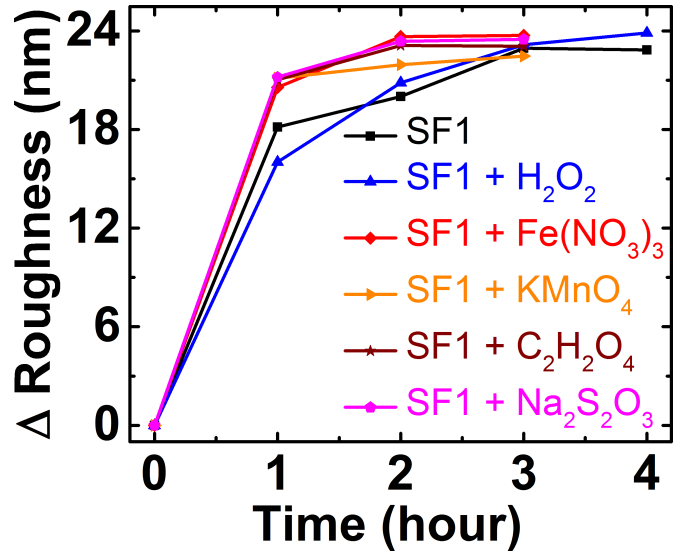}
\caption{Reduction in roughness as a function of polishing times. The slurries with solid additives reach their limiting values just after 2 hours of polishing, while the unmodified and \hp\ added slurry show no signs of plateauing off after 4 hours.} \label{figcomall}
\end{figure}

\begin{figure}
\centering
\includegraphics[width=0.7\textwidth]{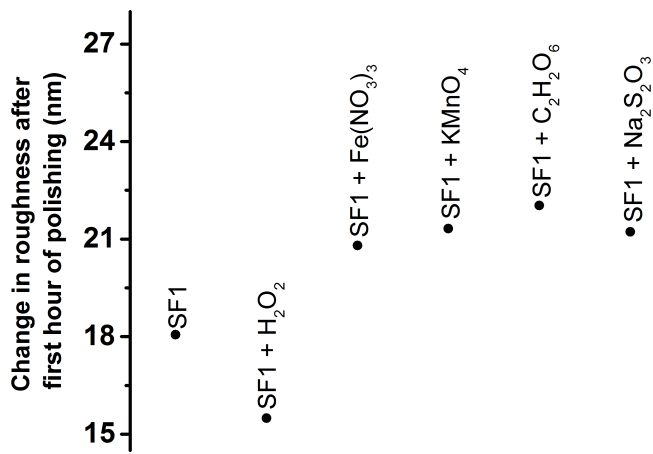}
\caption{Change in roughness after the first hour of polishing.  A clear reduction in roughness of almost 17\% is observable upon the addition of solid additives when compared to pure SF1.} \label{figcom}
\end{figure}

These results clearly show that the solid additives are accelerating the polishing process, all achieving a final roughness of under 2 nm RMS after 3 hours of polishing.  Hydrogen peroxide, even though it is a strong oxidising agent however appears to have very little effect on the polishing rate.  This may be due to the volatile nature of \hy\ which can easily escape from the mixture even at ambient temperature and pressure conditions.  Figure \ref{figcomall} shows the progression of the change in roughness at hourly intervals for the polished films, reiterating this clear difference between the slurries with solid additives and those without.  For the samples polished with the addition of \fn, \pp, \ox\, and \sd\ it can be seen that the roughness begins to plateau after 2 hours upon reaching a limit set by the granular nature of the film.  Those polished without additives or with \hy\ meanwhile continue to show a reduction in the roughness till 4 hours.  Figure \ref{figcom} then plots the change in roughness after 1 hour for the slurries.  While the addition of \hy{} to SF1 appears to initially worsens the polishing rate when comparing to pure SF1, a 17\% increase in reduction in roughness rate is observed upon the addition of solid additives.  Between the solid additives it can also be seen that little distinction in made between oxidising and reducing agents, with the reducing agent oxalic acid providing the fastest reduction in roughness rate.

\begin{figure}
\centering
\includegraphics[width=0.7\textwidth]{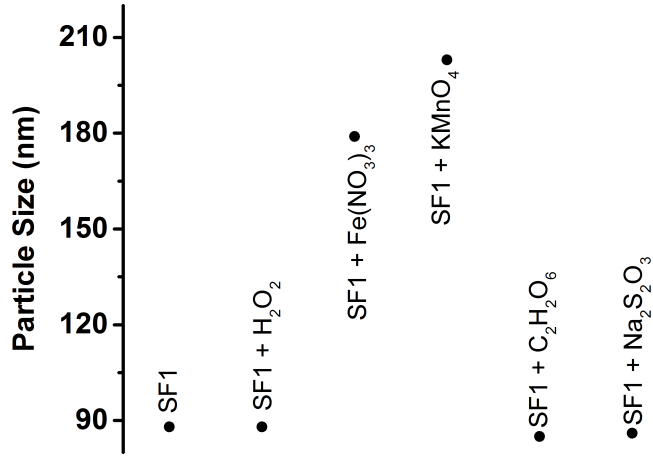}
\caption{Particle sizes of resultant SF1 slurries with various additives as measured by dynamic light scattering.  The addition of \fn\ and \pp\ result in an increase in aggregation of the 90 nm silica particles within the SF1 while the \hy, \ox\, and \sd\ additives leave the particle size unaltered.} \label{figpart}
\end{figure}

With earlier studies showing the reduction in roughness rate is proportional to the particle size, the size of slurry particles upon the addition of the redox agents was measured with the results shown within figure \ref{figpart} \cite{werrell2017}.  From the plot it can be seen that there is no change in particle size for the \hy, \ox{} and \sd{} additive slurries. This is a clear indication that the acceleration of polishing seen in the case of \ox{} or \sd{} is not due to change in particle size, rather it has more to do with the chemical nature of the slurry. The increase in size for \fn{} and \pp{} added slurries can be attributed to the agglomeration of the particles due to addition of the redox agents, which subsequently break apart upon mechanical action.  As such the variation in size does not correlate with the reduction in roughness rate.

Density function theory studies performed Peguiron et al.\cite{peguiron2016} have demonstrated that the polar silica molecule is able to break down the diamond surface due to weakening of C-C bonds between the terminating zigzag carbon chains.  According to their prediction, and confirmed by Werrell et al.\cite{werrell2017}, degradation of the diamond surface can be performed not only by silica, but also by alumina or ceria.  As such it is believed the observed acceleration in the reduction in roughness rate upon the addition of certain redox agents is due to acceleration in C-C bond breakage. Therefore, XPS has been performed on the polished samples to see whether the oxygen content on the surface differs from that seen on pure SF1 polished wafers \cite{thomas2014, werrell2017}.

\begin{figure}
\centering
\includegraphics[width=\textwidth]{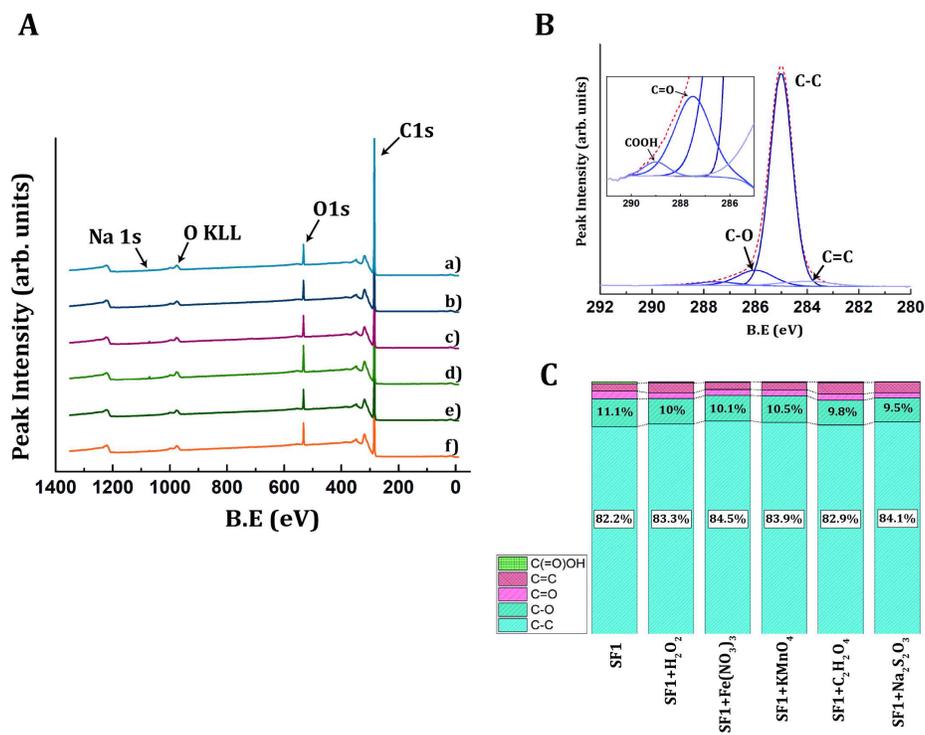}
\caption{A) Survey spectra of the polished films, B) deconvolution of the \oa\ C1s peak detailing the constituting environments, and C) constituent fractions of the C1s peak for the polished films.  Significant O1s character can be observed on all films along with traces of Na, Ca, and Si, while deconvolution demonstrates little change in the species terminating the NCD surface.} \label{xps}
\end{figure}

The resulting survey XPS spectra of the polished films are shown within panel A of Figure \ref{xps}, with the trace a) representing the SF1 polished sample and traces b)-f) representing those polished with the addition of \hp, \fen, \km, \oa, and \sod, respectively.  In accordance with previous studies, the polished films all possess significant C1s and O1s character at 285 eV and 533 eV, respectively \cite{thomas2014,werrell2017}.  Traces visible at 103 and 153 eV are meanwhile attributable to the 2s and 2p transitions of Si, indicating the presence of tightly bound slurry residues resistant to SC-1 cleaning, while further contamination can be observed with the appearance of peaks arising from Na and Ca.  In addition, the samples polished with the addition of \fen\ and \km\ show traces attributable to Fe and Mn, respectively, remaining after the cleaning process suggestive of chemical interaction with the NCD surface.  While not believed to be deleterious to the use of smooth NCD films, such contamination can likely be removed through streamlining of the SC-1 cleaning process before chemical absorption can occur \cite{liu2003}, or the use of other post-CMP cleaning techniques used within the IC-industry such as polyvinyl alcohol brush, buffered hydrofluoric acid, or remote hydrogen plasma \cite{zantye2004}.

To elucidate on the terminating species present at the surface, and hence the polishing mechanism, the C1s peak can be de-convoluted into its constituent components.  However, with different absolute values of binding energy (BE) often ascribed to the same functional groups and the small BE shifts between the components, precise determination of the surface configuration is often difficult \cite{ferro2005}.  BE shifts from the literature were therefore used to de-convolute the C1s line into the expected chemical environments.  As a result of the large inelastic mean free path of electrons originating from the C1s line, the main component of the peak at 285 eV is attributed to bulk and surface sp$^3$ C-C bonds, with a minor contribution from mono-hydrated surface dimers \cite{ghodbane2010}.  Further peaks at relative BE shifts of -1, +1, 2.5, and 4 eV attributable to graphitic carbon, ether/hydroxyl, carbonyl, and carboxyl groups, respectively, then provided an adequate fit to the data \cite{klauser2010}.  While CH$_x$ (x = 2,3) species are known to result in a component at 285.5 eV, the addition of a peak at a BE shift of +0.5 eV resulted in inconsistencies in the fitting procedure and predominantly altered the shape of the bulk sp$^3$ component, with marginal modification of the sp$^2$ and oxygen related peaks.  With similar behaviour noted in previous studies the component was then omitted from the fit \cite{klauser2010, ferro2005}.  Taking the \oa\ additive polished sample as an example, figure \ref{xps}B plots the C1s spectra and resulting peak components after deconvolution, while figure \ref{xps}C plots the components making up the C1s peak for the set of polished films.

\begin{table}
\centering
\begin{tabular}{ c c }
\hline
Slurry & O1s/C1s  \\ 
\hline
SF1 & 0.056\\
SF1 + \hp & 0.047\\
SF1 + \fen & 0.051\\
SF1 + \km & 0.061\\
SF1 + \oa & 0.051\\
SF1 + \sod & 0.046
\end{tabular}
\caption{O1s/C1s ratios for the as grown and 6 polished NCD samples.  Minor changes in the ratio occur for the 6 polished samples, with little correlation between the oxygen content and the slurries with the largest reduction in roughness or silica particle size, tentatively suggesting the oxygen content of the surface is not a determining factor in polishing.}
\label{O1sC1sratio}
\end{table}

Comparing the relative contents of the species composing the C1s peak within figure \ref{xps}C, little change in the ether/hydroxyl, carbonyl, and carboxyl fractions can be observed between the SF1 and modified slurry polished samples.  It is therefore unlikely that the difference in removal rate between the slurries can be attributed to a difference in the amount of oxygenated carbon present at the NCD surface.  In addition, little change in the graphitic bonded carbon peak occurs between the samples, suggesting that unlike traditional scaife based polishing, phase conversion to less dense forms of carbon is not responsible for the polishing observed \cite{peguiron2016}.  With little change in the species terminating the surface visible, Table \ref{O1sC1sratio} expresses the ratio of the area of the O1s peak with respect to the C1s peak.  In a similar fashion to previous studies the polished films all show larger ratios than that expected for an unpolished, hydrogenated film suggesting an increase in oxygen from silica or other contaminants \cite{thomas2014, werrell2017}.  Between the samples however there is little change in the ratios and no obvious correlation between the agents that accelerated the polishing rate and a larger O1s/C1s ratio.  This clearly indicates that the addition of redox agents do not increase the density of oxygen containing species on the surface.  The possible explanation to the increased polishing rate, in the event of mechanical process remaining the same, is the acceleration of the chemical reaction rate in the polishing mechanism i.e. the attachment of Si or O atom to the diamond surface.

\section{Conclusions}
In conclusion, a series diamond films were polished using SF1 slurry mixed with oxidizing agents, \hy, \pp, \fn{} and reducing agents, \ox, \sd{}. Oxalic acid addition led to the highest acceleration in the polishing rate of the diamond films as revealed by the AFM data. Particle size determination of the resulting slurry saw no evidence of particle size modification to effect the polishing rates. XPS studies carried out the polished film revealed non-significant variation of oxygen content on the diamond surface. These results demonstrated a probable rate rise in the chemical side of the process thus accelerating the polishing process.

\section*{Acknowledgments}
The authors wish to acknowledge the financial support of the European Research Council (ERC) Consolidator Grant for the development of ‘Superconducting Diamond Quantum Nano-Electro-Mechanical Systems’, Project ID: 647471.

\section*{References}

\bibliography{mybibfile}

\end{document}